\documentclass[11pt,a4paper]{article}
\usepackage{jcappub}
\usepackage{euscript,epsfig,amsmath,amssymb}
\usepackage{amsfonts,latexsym}

\newcommand{\be}[1]{\begin{equation}\label{#1}}
\newcommand{\ee}{\end{equation}}
\newcommand{\ba}[1]{\begin{eqnarray}\label{#1}}
\newcommand{\ea}{\end{eqnarray}}
\newcommand{\rf}[1]{(\ref{#1})}
\newcommand{\nn}{\nonumber}
\renewcommand{\theequation}{\arabic{section}.\arabic{equation}}

\newcommand{\mc}{\mathcal}

\newcommand{\de}{\partial}
\newcommand{\al}{\alpha}
\newcommand{\la}{\lambda}

\newcommand{\hg}{\hat g}
\newcommand{\hga}{\hat\Gamma}
\newcommand{\hna}{\hat\nabla}
\newcommand{\Ga}{\Gamma}
\newcommand{\hr}{\hat R}

\begin{document}

\title{Kaluza-Klein multidimensional models with Ricci-flat internal spaces:\\ the absence of the KK particles}

\author{Alexey Chopovsky$^{1,2}$,}
\author{Maxim Eingorn$^{3}$,}
\author{and Alexander Zhuk$^{2}$}

\affiliation{$^{1}$Department of Theoretical Physics, Odessa National University,\\ Dvoryanskaya st. 2, Odessa 65082, Ukraine\\}

\affiliation{$^{2}$Astronomical Observatory, Odessa National University,\\ Dvoryanskaya st. 2, Odessa 65082, Ukraine\\}

\affiliation{$^{3}$North Carolina Central University, CREST and NASA Research Centers,\\ Fayetteville st. 1801, Durham, North Carolina 27707, U.S.A.\\}

\emailAdd{a.chopovsky@yandex.ru} \emailAdd{maxim.eingorn@gmail.com} \emailAdd{ai.zhuk2@gmail.com}

\abstract{In this paper we consider a multidimensional Kaluza-Klein (KK) model with a Ricci-flat internal space, e.g., a Calabi-Yau manifold. We perturb this
background metrics by a system of gravitating masses, e.g., astrophysical objects such as our Sun. We suppose that these masses are pressureless in the
external space but they have relativistic pressure in the internal space. We show that metric perturbations do not depend on coordinates of the internal space
and gravitating masses should be uniformly smeared over the internal space. This means, first, that KK modes corresponding to the metric fluctuations are
absent and, second, particles should be only in the ground quantum state with respect to the internal space. In our opinion, these results look very unnatural.
According to statistical physics, any nonzero temperature should result in fluctuations, i.e. in KK modes. We also get formulae for the metric correction terms
which enable to calculate the gravitational tests: the deflection of light, the time-delay of the radar echoes and the perihelion advance.}

\maketitle

\flushbottom

\section{\label{sec:1}Introduction}

It may turn out that our Universe is multidimensional and has more than three spatial dimensions. This fantastic idea was first proposed in the twenties of the
last century in the papers by T. Kaluza and O. Klein \cite{Kaluza,Klein}, and since then it has consistently attracted attention of researchers. Obviously, to
be compatible with our observable four-dimensional spacetime, the extra dimensions should be compact and sufficiently small. According to the recent
accelerator experiments, their size should be of the order or less than the Fermi length $10^{-17}$ cm\footnote{In the brane-world models, the extra dimensions
can be significantly larger \cite{ADD,AADD}. We do not consider such models in our paper.}. If the internal space is very small, the hope to detect them is
connected with Kaluza-Klein (KK) particles which correspond to appropriate eigenfunctions of the internal manifold, i.e. excitations of fields in the given
internal space. Such excitations were investigated in many articles (see, e.g., the classical papers \cite{SS,Nieu,Kim}). Quite recently, KK particles were
considered, e.g., in the papers \cite{KLZ,BD}.

The internal compact space can have different topology. The Ricci-flat space is of special interest because Calabi-Yau manifolds, which are widely used in the
superstring theory \cite{Superstrings}, belong to this class. A particular case of toroidal compactification was considered in our previous paper \cite{ChEZ5}.
We demonstrate in this article that metric perturbations depend only on coordinates of the external spacetime, i.e. the corresponding KK particles are absent.
In our approach, the metric perturbations are caused by a system of gravitating masses, e.g., by astrophysical objects. This is the main difference between our
investigation and studies in \cite{KLZ,BD} where the metric and form-field perturbations are considered without taking into account the reason of such
fluctuations. Our analysis in \cite{ChEZ5} clearly shows that the inclusion of the matter sources, being responsible for the perturbations, imposes strong
restrictions on the model. In the present paper, we investigate this important phenomenon in the case of Ricci-flat internal spaces and arrive again at the
same conclusions. That is, KK modes corresponding to the metric fluctuations are absent. We also show that gravitating masses should be uniformly smeared over
the internal space. From the point of quantum mechanics, it means that particles must be only in the ground state with respect to the internal space. Of
course, it looks very unnatural. According to statistical physics, any nonzero temperature should result in fluctuations, i.e. in KK states.

The paper is structured as follows. In Sec. 2, we consider the multidimensional model with the Ricci-flat internal space. This background spacetime is
perturbed by the system of gravitating masses. We demonstrate that the metric perturbations do not depend on the coordinates of the internal space and
gravitating masses are uniformly smeared over the internal space. That is, KK modes are absent. In Sec. 3, we provide the exact expressions for the metric
correction terms. The main results are briefly summarized in concluding Sec. 4.


\section{\label{sec:2}Smeared extra dimensions}

We consider a block-diagonal background metrics
\be{2.1}
[\hat
g^{(\mc{D})}_{MN}(y)]=[\eta^{(4)}_{\mu\nu}]\oplus[\hat g^{(d)}_{mn}(y)]\, ,
\ee
which is determined on $(\mathcal{D}=1+D=4+d)-$dimensional product manifold $\mathfrak{M}_{\mathcal D}=\mathfrak{M}_4\times \mathfrak{M}_{d}$. Here,
$\eta^{(4)}_{\mu\nu}\, (\mu,\nu = 0,1,2,3)$ is the metrics of the four-dimensional Minkowski spacetime, and $\hat g^{(d)}_{mn}(y)\, (m,n=4,\ldots ,D)$ is the
metrics of some $d$-dimensional internal (usually, compact) Ricci-flat space with coordinates $y\equiv y^m$:
\be{2.2} \hat R^{(d)}_{mn}[\hat g^{(d)}_{mn}]=0\, . \ee
In what follows, $X^M \, (M=0,1,2,\ldots ,D)$ are coordinates on $\mathfrak{M}_{\mathcal{D}}$, $x^{\mu}=X^{\mu}\, (\mu = 0,1,2,3)$ are coordinates on
$\mathfrak{M}_4$ and $y^{m}=X^m\, (m =4,5,\ldots ,3+d)$ are coordinates on $\mathfrak{M}_d$. We will also use the definitions for the spatial coordinates
$\tilde X:\; X^{\tilde M}\, (\tilde M = 1,2,\ldots ,D)$ and $\tilde x:\; x^{\tilde \mu} = X^{\tilde \mu}\, (\tilde \mu = 1,2,3)$.

Now, we perturb the metrics \rf{2.1} by a system of $N$ discrete massive (with rest masses $m_p,\, p=1,\ldots ,N$) bodies. We suppose that the pressure of
these bodies in the external three-dimensional space is much less than their energy density. This is a natural approximation for ordinary astrophysical objects
such as our Sun. For example, in general relativity, this approach works well for calculating the gravitational experiments in the Solar system \cite{Landau}.
In other words, the gravitating bodies are pressureless in the external/our space. On the other hand, we suppose that they may have pressure in the extra
dimensions. Therefore, nonzero components of the energy-momentum tensor of the system can be written in the form for a perfect fluid \cite{Landau,ChEZ5}:
\begin{eqnarray}\label{2.3}
  T^{M \nu}&=&\rho c^2 \frac{ds}
{dx^0} u^M u^\nu,\quad u^M=\frac{dX^M}{ds},\nonumber\\
  T^{mn}&=&-p_mg^{mn}+\rho c^2 \frac{ds}
{dx^0} u^m u^n\, ,
\end{eqnarray}
with the equation of state (EoS) in the internal space
\be{2.4} p_m=\omega\rho c^2\frac{ds}{dx^0}\, ,\quad m=4,5,\ldots ,D\, , \ee
where $\omega$ is the parameter of EoS{\footnote{We should note that the pressure \rf{2.4} is the intrinsic pressure of a gravitating mass in the extra
dimensions and is not connected with its motion. In other words, $\omega$ is the parameter of a particle.}}  and the rest-mass density is
\be{2.5} \rho \equiv \sum_{p=1}^N [|g^{\mathcal{(D)}}|]^{-1/2}m_p\delta({\bf \tilde X}-{\bf \tilde X}_p)\, , \ee
where ${\bf \tilde X}_p$ is a $D$-dimensional "radius-vector" of the $p-$th particle. In EoS, we include the factor $ds/dx^0$ just for convenience. We can also
eliminate this factor from EoS and include it in the definition of the rest-mass density \rf{2.5} (see, e.g., \cite{ChEZ5}). For our purpose, it is sufficient
to consider a steady-state model{\footnote{For example, we can consider just one gravitating mass and place the origin of a reference frame on it.}}. That is,
we disregard the spatial velocities of the gravitating masses.

In the weak-field approximation, the perturbed metric components can be written in the following form:
\be{2.6}
g_{MN}\approx\hat g_{MN}+h_{MN}, \quad g^{MN}\approx \hat g^{MN}-h^{MN},
\ee
%
where $h_{MN}\sim O(1/c^2)$ and $h_M^N\equiv \hat g^{NT}h_{MT}$. Hereafter, the "hats" denote the unperturbed quantities. For example, the metric coefficients
in \rf{2.3}-\rf{2.5} are already perturbed quantities.

To get the metric correction terms, we should solve (in the corresponding order) the multidimensional Einstein equation
\be{2.7}
R_{MN}=\cfrac{2S_{D}\tilde G_{\mc D}}{c^4}\left[T_{MN}-\cfrac{1}{D-1}\,Tg_{MN}\right]\, ,
\ee
where $S_D=2\pi^{D/2}/\Gamma (D/2)$ is the total solid angle (the surface area of the $(D-1)$-dimensional sphere of the unit radius), $\tilde G_{\mathcal{D}}$
is the gravitational constant in the $(\mathcal{D}=D+1)$-dimensional spacetime. For our model and up to the $O(1/c^2)$ terms, the Einstein equation \rf{2.7} is
reduced to the following system of equations (see \cite{ChEZ5} for details):
\ba{2.8}
R_{00}[g^{(\mc D)}]&\approx & -\cfrac{1}{2}\,\hna_{\tilde L}\hna^{\tilde L}h_{00}=
\cfrac{2S_D\tilde G_{\mathcal D}}{c^2}\,\left(\cfrac{D-2+\omega d}{D-1}\right)\,\rho\, ,\quad \tilde L = 1,2,\ldots ,D\, ,\\
\label{2.9}
R_{\tilde\mu\tilde\nu}[g^{(\mc D)}]&\approx & -\cfrac{1}{2}\,\hna_{\tilde L}\hna^{\tilde L}h_{\tilde\mu\tilde\nu}=\cfrac{2S_D\tilde
G_{\mathcal D}}{c^2}\,\left(\cfrac{1-\omega d}{D-1}\right)\,\rho\delta_{\tilde\mu\tilde\nu}\, ,\quad \tilde\mu,\tilde\nu =1,2,3 \, ,\\
\label{2.10} R_{mn}[g^{(\mc D)}]&\approx & -\cfrac{1}{2}\,\hna_{\tilde L}\hna^{\tilde L}h_{mn}=-\cfrac{2S_D\tilde G_{\mathcal
D}}{c^2}\,\left(\cfrac{\omega(D-1)+1-\omega d}{D-1}\right)\,\rho \hg_{mn}\, , \ea
where the Ricci-tensor components $R_{MN}[g^{(\mc D)}]$ are determined by the equation \rf{a12}.

Obviously, the solution of this system is determined by the rest-mass density $\rho (\tilde X)$. Let us suppose, e.g., that the physically reasonable solution
$h_{00} (\tilde X)$ of \rf{2.8} exists\footnote{The exact form of such solution depends on the topology of the internal space. For example, in the case of
toroidal internal spaces with the periodic boundary conditions in the directions of the extra dimensions, the corresponding solutions can be found in
\cite{EZ1,EZ2}. However, as we will see later, the exact form of $D$-dimensional solutions is not important for our purpose. A similar situation took place in
our article \cite{EZ3} where this point was discussed in detail after Eq. (2.56).}. Then, we can write it in the form
\be{2.11}
h_{00}(\tilde X)\equiv 2\varphi (\tilde X)/c^2\, .
\ee
It is clear that the function $\varphi (\tilde X)$ describes the nonrelativistic gravitational potential created by the system of masses with the rest-mass
density \rf{2.5}. From Eqs. \rf{2.9} and \rf{2.10} we get
\be{2.12} h_{\tilde\mu\tilde\nu}=\cfrac{1-\omega d}{D-2+\omega d}\,h_{00}\delta_{\tilde\mu\tilde\nu} \ee
and
\be{2.13} h_{mn}=-\cfrac{\omega(D-1)+1-\omega d}{D-2+\omega d}\,h_{00}\hg_{mn}\, . \ee
It is worth noting that now we can easily determine (via the gravitational potential $\varphi (\tilde X)$) the functions $\xi_1,\xi_2$ and $\xi_3$ introduced
in \rf{a11}.

It is well known (see, e.g., \cite{EZ6}) that to be in agreement with the gravitational tests (the deflection of light and the Shapiro time-delay experiment)
at the same level of accuracy as general relativity, the metric correction terms should satisfy the equalities
\be{2.14}
h_{00}=h_{11}=h_{22}=h_{33}\, .
\ee
This results in the black string/brane condition \cite{EZ4,EZ5}
\be{2.15}
\omega =-\frac12\, .
\ee
Thus, to be in agreement with observations, this parameter should not be negligibly small, i.e. $\omega \sim O(1)$.

The solutions \rf{2.11}-\rf{2.13} should satisfy the gauge condition \rf{a7}. Taking into account that $h = \hg^{MN}h_{MN}=[2(\omega d-1)/(D-2+\omega
d)]\,h_{00}$, we can easily get that for $N=\nu\, \ (\nu =0,1,2,3)$ this condition is automatically satisfied. However, in the case $N=n\, \ (n=4,5,\ldots ,D)$
we arrive at the condition
\be{2.16} \hna_{L}\hg^{LT}h_{Tn}-\cfrac{1}{2}\,\de_n h=-\cfrac{\omega(D-1)}{D-2+\omega d}\,\de_nh_{00}=0\, . \ee
Because $\omega \neq 0$, to satisfy this equation, we must demand that
\be{2.17}
\de_n h_{00}=\cfrac{2}{c^2}\,\de_n\varphi(\tilde X)=0\, .
\ee
Hence, the gravitational potential is the function of the spatial coordinates of the external/our space: $\varphi(\tilde X)=\varphi(\tilde x)$. Obviously, it
takes place only if the rest-mass density also depends only on the spatial coordinates of the external space: $\rho(\tilde X)=\rho(\tilde x)$. In other words,
the gravitating masses should be uniformly smeared over the internal space. For short, we usually call such internal spaces "smeared" extra dimensions.

This conclusion has the following important effect. Suppose that we have solved for the considered particle a multidimensional quantum
Schr$\mathrm{\ddot{o}}$dinger equation and found its wave function $\Psi (\tilde X)$. In general, this function depends on all spatial coordinates $\tilde
X=(\tilde x,y)$, and we can expand it in appropriate eigenfunctions of the compact internal space, i.e. in the Kaluza-Klein (KK) modes. The ground state
corresponds to the absence of these particles. In this state the wave function may depend only on the coordinates $\tilde x$ of the external space. The
classical rest-mass density is proportional to the probability density $|\Psi|^2$. Therefore, the demand, that the rest-mass density depends only on the
coordinates of the external space, means that the particle can be only in the ground quantum state, and KK excitations are absent. This looks very unnatural
from the point of quantum and statistical physics.



\section{\label{sec:3}Metric components}

Now, we want to get the expression for the metrics up to $1/c^2$ terms. Since $x^0=ct$, the component $g_{00}$ should be calculated up to $O(1/c^4)$:
\be{3.1} g_{00}\approx \eta_{00}+h_{00}+f_{00}\, ,\ee
where $h_{00}\sim O(1/c^2)$ and $f_{00}\sim O(1/c^4)$. For the rest metric coefficients, we still have Eq. \rf{2.6}: $g_{\tilde M\tilde N}\approx\hat g_{\tilde
M\tilde N}+h_{\tilde M\tilde N}\, ,\, h_{\tilde M\tilde N}\sim O(1/c^2)$. We remind that we consider the steady-state case: $g_{0\tilde M}=0$.

According to the results of the previous section, the gravitating masses are uniformly smeared over the internal space. Therefore, the rest-mass density
\rf{2.5} now reads
\be{3.2} \rho=\cfrac{1}{V_d}\sum_{p=1}^Nm_p\cfrac{\delta(\tilde {\bf x}-\tilde {\bf x}_p)}{\sqrt{-g^{(4)}}}\, =
\cfrac{1}{V_d}\cfrac{\varrho_3}{\sqrt{-g^{(4)}}}\, , \ee
where $V_d=\int_{{\mathfrak M}_d}d^dy\sqrt{|g^{(d)}|}$ is the volume of the internal space and
\be{3.3}
\varrho_3 (\tilde x) \equiv \sum_{p=1}^N m_p \delta(\tilde {\bf x}-\tilde {\bf x}_p)\, .
\ee
Here, $\tilde {\bf x}$ is a three-dimensional radius-vector in the external space. Additionally, it is clear that all prefactors $\xi_1,\xi_2$ and $\xi_3$ in
\rf{a11} are also functions of the external spatial coordinates $\tilde x$. For convenience, we redetermine the function $\xi_3$: $\xi_3 (\tilde x)\equiv
\alpha (\tilde x)$. Then, $h_{mn}(x, y)=\alpha(x)\hat{g}_{mn}(y)$, where $\alpha(x)\sim O(1/c^2)$.

To get the metric correction terms, we should solve the Einstein equation \rf{2.7}. First, we consider the $O(1/c^2)$ terms $h_{MN}$. These terms satisfy the
system \rf{2.8}-\rf{2.10} where we must perform the following replacements: $\rho \, \rightarrow \, \varrho_3/\hat V_d$, where $\hat V_d=\int_{{\mathfrak
M}_d}d^dy\sqrt{|\hat g^{(d)}|}$ is the volume of the unperturbed internal space, and $-\hna_{\tilde L}\hna^{\tilde L} \, \rightarrow\, \triangle =
\delta^{\tilde \mu \tilde \nu}\partial^2/\partial x^{\tilde\mu}\partial x^{\tilde\nu}$. Introducing the definition $h_{00}\equiv 2\varphi/c^2$, we can easily
get from this system
\be{3.4} \varphi=-G_N\sum_{p=1}^N\cfrac{m_p}{|\tilde {\bf x}-\tilde {\bf x}_p|}\, , \ee
where
\be{3.5} G_N=\cfrac{S_D (D-2+\omega d)}{2\pi \hat V_d (D-1)}\,\tilde G_{\mc D} \ee
is the Newtonian gravitational constant and $\varphi (\tilde x)$ is the nonrelativistic gravitational potential created by $N$ masses. At the same time
\ba{3.6}
h_{\tilde\mu\tilde\nu}&=&\cfrac{1-\omega d}{D-2+\omega d}\,h_{00}\delta_{\tilde\mu\tilde\nu}\, ,\\
\label{3.7} h_{mn}(x, y)&=&\alpha(x)\hat{g}_{mn}(y)\, ,\quad \alpha=-\cfrac{\omega(D-1)+1-\omega d}{D-2+\omega d}\, h_{00}\, . \ea

Now, we intend to determine the $f_{00}\sim O(1/c^4)$ correction term. Obviously, to do it, we need to solve only the 00-component of the Einstein equation
\rf{2.7}. Keeping in mind the prefactor $1/c^4$ in the right hand side of \rf{2.7}, it is clear that the energy-momentum tensor component $T_{00}$ and the
trace $T$ must be determined up to $O(1)$. Therefore, similar to the paper \cite{ChEZ5}, we get:
\ba{3.8}
T_{00}&\approx&\cfrac{\varrho_3 c^2}{\hat V_d}\left[1+\cfrac{3D-4+\omega d}{2(D-2+\omega d)}\,h_{00} \right]\, ,\\
\label{3.9} T&\approx&\cfrac{\varrho_3 c^2}{\hat V_d}\left[1+\cfrac{D-\omega d}{2(D-2+\omega d)}\,h_{00} \right]\left(1-\omega d\right)\, . \ea
To get the Ricci-tensor component $R_{00}$ up to $O(1/c^4)$, we can use Eq. \rf{b2} and Eqs. (2.29), (2.38), (2.39) in \cite{EZ3} and arrive at the following
formula:
\be{3.10} R_{00}\approx \cfrac{1}{2}\,\triangle\left(f_{00}-\cfrac{2}{c^4}\,\varphi^2\right)+\cfrac{1}{c^2}\,\triangle\varphi +\cfrac{D-1}{D-2+\omega
d}\,\cfrac{2}{c^4}\,\varphi\triangle\varphi\, , \ee
where $\triangle = \delta^{\tilde \mu \tilde \nu}\partial^2/\partial x^{\tilde\mu}\partial x^{\tilde\nu}$ is the 3-dimensional Laplace operator. Therefore, the
$00$-component of the Einstein equation reads
\be{3.11} \cfrac{1}{2}\,\triangle\left(f_{00}-\cfrac{2}{c^4}\,\varphi^2\right)=\cfrac{1}{c^4}\,\varphi\triangle\varphi =\frac{4\pi G_N}{c^4}\sum_{p=1}^N
m_p\varphi^{\prime}(\tilde {\bf x}_p)\delta(\tilde {\bf x} -\tilde {\bf x}_p)\, , \ee
where we used the equation $\triangle \varphi = 4\pi G_N \sum_{p=1}^N m_p \delta(\tilde {\bf x} -\tilde {\bf x}_p)$. The function $\varphi'(\tilde {\bf x}_p)$
is the potential of the gravitational field at a point with the radius vector ${\bf x}_p$ produced by all particles, except the $p$-th:
\be{3.12}
\varphi' (\tilde {\bf x}_p) = \sum_{q\neq p} \varphi_q (\tilde {\bf x}_p-\tilde {\bf x}_q)
\ee
and
\be{3.13}
\triangle \varphi_p (\tilde {\bf x}-\tilde {\bf x}_p) = 4\pi G_N m_p \delta(\tilde {\bf x} -\tilde {\bf x}_p)\, .
\ee
Therefore, the solution of Eq. \rf{3.11} is
\be{3.14} f_{00}(\tilde {\bf x})=\cfrac{2}{c^4}\,\varphi^2(\tilde {\bf x}) +\cfrac{2}{c^4}\sum_{p=1}^N\varphi_p(\tilde {\bf x}-\tilde {\bf x}_p)\varphi'(\tilde
{\bf x}_p)\, . \ee
Obviously, in the case of the toroidal internal space, this section reproduces the results of our paper \cite{ChEZ5}. The found metric correction terms enable
to calculate the gravitational tests (the deflection of light, the time-delay of the radar echoes and the perihelion advance) in full analogy with the paper
\cite{EZ5}.


\section{Conclusion}

In our paper, we have studied the multidimensional KK model with the Ricci-flat internal space. Such models are of special interest because Calabi-Yau
manifolds, which are widely used in the superstring theory \cite{Superstrings},  belong to this class. In the absence of matter sources, the background
manifold is a direct product of the four-dimensional Minkowski spacetime and the Ricci-flat internal space. Then, we perturbed this background metrics by a
system of gravitating masses, e.g., astrophysical objects such as our Sun. It is well known that pressure inside of these objects is much less than the energy
density. Therefore, we supposed that gravitating masses are pressureless in the external space. However, they may have relativistic pressure in the internal
space. Moreover, the presence of such pressure is the necessary condition for agreement with the gravitational tests (the deflection of light, the time-delay
of the radar echoes and the perihelion advance) \cite{EZ6,EZ4,EZ5}. We clearly demonstrated that metric perturbations do not depend on the coordinates of the
internal space and gravitating masses are uniformly smeared over the internal space. This means, first, that KK modes corresponding to the metric fluctuations
are absent and, second, the particles corresponding to the gravitating masses should be only in the ground quantum state with respect to the internal space. Of
course, it looks very unnatural. According to statistical physics, any nonzero temperature should result in fluctuations, i.e. in KK modes. On the other hand,
the conclusion about the absence of KK particles is consistent with the results of \cite{KKR,BR}, where a mechanism  for the symmetries formation of the
internal spaces was proposed.  According to this mechanism, due to the entropy decreasing in the compact subspace, its metric undergoes the process of
symmetrization during some time after its quantum nucleation. A high symmetry of internal spaces actually means suppression of the extra degrees of freedom.

We also obtained the formulae for the metric correction terms. For the 00-component, we got it up to $O(1/c^4)$. We found other components up to $O(1/c^2)$.
This enables to calculate the gravitational tests: the deflection of light, the time-delay of the radar echoes and the perihelion advance. It also gives a
possibility to construct the Lagrange function for the considered many-body system (see, e.g., \cite{ChEZ5}).


\section*{Acknowledgements}

This work was supported by the "Cosmomicrophysics-2" programme of the Physics and Astronomy Division of the National Academy of Sciences of Ukraine. The work
of M. Eingorn was supported by NSF CREST award HRD-0833184 and NASA grant NNX09AV07A.


\section*{Appendix A: Ricci tensor in the weak-field approximation}
\renewcommand{\theequation}{A.\arabic{equation}}
\setcounter{equation}{0}

In our paper we use the signature of the metrics $(+,-,-,-,\ldots)$ and determine the Ricci tensor components as follows:
\be{a1}
R_{MN}\equiv R^S_{\,\,MSN}=\de_L\Ga^L_{MN}-\de_N\Ga^L_{ML}+\Ga^L_{MN}\Ga^S_{LS}-\Ga^S_{ML}\Ga^L_{NS}\, ,
\ee
where the Christoffel symbols are
\be{a2}
\Ga^L_{MN}=\cfrac{1}{2}\,g^{LT}\left(-\de_T g_{MN}+\de_M g_{NT}+\de_N g_{MT}\right)\, ,\quad M,N,L, \ldots = 0,1,2,\ldots ,D\, .
\ee
The metrics $g^{(\mathcal{D})}_{MN}\equiv g_{MN}$ is determined on $(\mathcal{D}=1+D)$-dimensional manifold $\mathfrak{M}_{\mathcal{D}}$.

Now, we suppose that the metric coefficients $g_{MN}$ correspond to the perturbed metrics, and in the weak-field approximation they can be decomposed as
\be{a3}
g_{MN}\approx\hat g_{MN}+h_{MN}, \quad g^{MN}\approx \hat g^{MN}-h^{MN}\, ,
\ee
where $\hat g_{MN}$ is the background unperturbed metrics and the metric perturbations $h_{MN}\sim O(1/c^2)$ ($c$ is the speed of light). As usual,
$h_M^N\equiv \hat g^{NT}h_{TM}$. Here and after, the "hats" denote unperturbed quantities. Then, up to linear terms $h$, the Christoffel symbols read
\begin{eqnarray}\label{a4}
\Ga^L_{MN}&\approx&\hga^L_{MN}+\cfrac{1}{2}\,\hg^{LT}\left[ -\de_Th_{MN}+\left(\de_M h_{NT}-\hga^P_{MN}h_{PT}\right)+
\left( \de_Nh_{MT}-\hga^P_{NM}h_{PT} \right) \right] \nn \\
&=&\hga^L_{MN}+\cfrac{1}{2}\,\hg^{LT}\left[-\hna_T h_{MN}+\hna_M h_{NT}+\hna_N h_{MT}\right]\, ,
\end{eqnarray}
and for the Ricci tensor we get
\be{a5}
R_{MN} \approx \hr_{MN}+\cfrac{1}{2}\left(-\hna_L\hna^Lh_{MN}+ Q_{MN}\right)\, ,
\ee
where we introduced a new tensor
\ba{a6}
Q_{MN} &\equiv& \hna_L\hna_Mh^L_N+\hna_L\hna_Nh^L_M-\hna_N\hna_Mh^L_L\nonumber\\
&=&
\left[\hna_M\left(\hna_Lh^L_N-\cfrac{1}{2}\,\de_Nh_L^L\right)+\hna_N\left(\hna_Lh^L_M-\cfrac{1}{2}\,\de_Mh_L^L\right)\right]\nonumber\\
&&-\left(\hr^L_{\,\,NPM}+\hr^L_{\,\,MPN}\right)h^P_L+\hr_{PM}h^P_N+\hr_{PN}h^P_M\, .
\ea
To get the last expression, we used the relation (91.8) in \cite{Landau}. The gauge conditions (see, e.g., the paragraph 105 in \cite{Landau})
\be{a7} \hna_Lh^L_N-\cfrac{1}{2}\,\de_Nh^L_L=0 \ee
simplify considerably this expression:
\be{a8}
Q_{MN}=-\left(\hr^L_{\,\,NPM}+\hr^L_{\,\,MPN}\right)h^P_L+\hr_{PM}h^P_N+\hr_{PN}h^P_M\, .
\ee

Now, we suppose that the manifold $\mathfrak{M}_{\mathcal D}$ is a product manifold $\mathfrak{M}_{\mathcal D}=\mathfrak{M}_4\times \mathfrak{M}_{d}$ with a
block-diagonal background metrics of the form \rf{2.1}. However, in this appendix we do not assume that the internal manifold $\mathfrak{M}_{d}$ is a
Ricci-flat one. For the metrics \rf{2.1} and an arbitrary internal space $\mathfrak{M}_d$, the only nonzero components of the Riemann tensor $\hat
R_{LNMP}[\hat g^{(\mc D)}]$ and the Ricci tensor $\hat R_{NM}[\hat g^{(\mc D)}]$ are
\be{a10} \hat R_{lnmp}[\hat g^{(\mc D)}]=\hat R_{lnmp}[\hat g^{(d)}], \quad \hat R_{mn}[\hat g^{(\mc D)}]=\hat R_{mn}[\hat g^{(d)}]\, . \ee

Typically, the matter, which perturbs the background metrics, is taken in the form of a perfect fluid with isotropic pressure. In the case of the product
manifold, the pressure is isotropic in each factor manifolds (see, e.g., the energy-momentum tensor \rf{2.3}). Obviously, such perturbation does not change the
topology of the model, e.g., the topologies of the factor manifolds in the case of the product manifold. Additionally, it preserves also the block-diagonal
structure of the metric tensor. In the case of a steady-state model (our case) the non-diagonal perturbations $h_{0{\tilde M}}\, (\tilde M = 1,2,3,\ldots ,D)$
are also absent. Therefore, the metric correction terms are conformal to the background metrics:
\ba{a11}
h_{00} &=&\xi_1 \hat g_{00}^{(\mathcal{D})}=\xi_1 \eta_{00}\, ,\quad
h_{\tilde\mu\tilde\nu}=\xi_2 \hat g_{\tilde\mu\tilde\nu}^{(\mathcal{D})}=-\xi_2\delta_{\tilde\mu\tilde\nu}\, , \ \tilde\mu,\tilde\nu = 1,2,3\, , \nn\\
h_{mn}&=& \xi_3 \hat g_{mn}^{(\mathcal{D})}= \xi_3 \hat g_{mn}^{(d)}\, , \ea
where $\xi_{1, 2, 3}$ are some scalar functions of all spatial coordinates $\tilde X$. It can be easily verified that under these conditions{\footnote{It is
worth noting that, as it follows from Eq. \rf{a8}, $Q_{MN}$ is also equal to zero when the following condition holds: $\hr^{L}_{\,\,NPM}=-\hr^{L}_{\,\,MPN}$,
resulting in the Ricci-flat space $\hat R_{MN}=0$ (but not vice versa).}} the tensor $Q_{MN}$ in Eq. \rf{a8} is equal to zero: $Q_{MN}=0$. Therefore, up to
$O(1/c^2)$ terms, we get
\be{a12} R_{MN}[g^{(\mc D)}]\approx \hat R_{MN}[\hat g^{(\mc D)}] -\cfrac{1}{2}\,\hna_L\hna^Lh_{MN}= \hat R_{MN}[\hat g^{(\mc D)}]-\cfrac{1}{2}\,\hna_{\tilde
L}\hna^{\tilde L}h_{MN} +o(1/c^2)\, , \ee
where $\tilde L = 1,2,\ldots ,D$ and we took into account that $X^0 = ct$.


\section*{Appendix B: Ricci-tensor for a block-diagonal metrics}
\renewcommand{\theequation}{B.\arabic{equation}}
\setcounter{equation}{0}

In this appendix we consider a block-diagonal metrics
\be{b1}
[g^{(\mathcal{D})}_{MN}(x, y)]=[g^{(d_1)}_{\mu\nu}(x)]\oplus[g^{(d_2)}_{mn}(x, y)]=[g^{(d_1)}_{\mu\nu}(x)]\oplus[e^{\al(x)}\hat g^{(d_2)}_{mn}(y)]\, .
\ee
Then, for the Ricci-tensor components we get the following expressions
\ba{b2}
R_{\mu\nu}[g_{MN}]&=&
R_{\mu\nu}[g^{(d_1)}_{\lambda\tau}]-\cfrac{d_2}{2}\left[\nabla^{(d_1)}_{\mu}(\de_\nu \al)+\cfrac{1}{2}\,(\de_\mu \al)(\de_\nu \al)\right]\, ,\\
\label{b3}
R_{m\nu}[g_{MN}]&=& 0\, ,\\
\label{b4} R_{m n}[g_{MN}]&=& R_{mn}[g^{(d_2)}_{kl}]-\cfrac{1}{2}e^{\al(x)}\,\hat g^{(d_2)}_{mn}g^{(d_1)\la\tau}\left[ \nabla^{(d_1)}_\la(\de_\tau \al)+
\cfrac{d_2}{2}\,(\de_\la \al)(\de_\tau \al)\right]\, . \ea

\end{document}